\documentclass[12pt,preprint]{aastex}
\shorttitle{Interaction of ejecta with an ambient medium}
\shortauthors{Nakamura \& Shigeyama}

\def\Msun{~M_{\odot} }

\begin{document}

\title{Self-similar solutions for the interaction of relativistic ejecta with an ambient medium}

\author{Ko Nakamura$^{1,2}$
\and Toshikazu Shigeyama$^1$}
\affil{$^1$Research Center for the Early Universe, Graduate School of Science, 
        University of Tokyo, Bunkyo-ku, Tokyo 113-0033, Japan\\
        $^2$Department of Astronomy, Graduate School of Science, 
        University of Tokyo, Bunkyo-ku, Tokyo 113-0033, Japan}

\begin{abstract}
We find self-similar solutions to describe the interaction of spherically symmetric ejecta expanding at relativistic speeds with an ambient medium having a power law density distribution.  Using this solution, the time evolution of the Lorentz factor of the outer shock is derived as a function of the explosion energy, the mass of the ejecta, and  parameters for the ambient medium. These solutions are an ultra-relativistic version of the solutions for the circumstellar interaction of supernova ejecta obtained by Chevalier and extensions of the relativistic blast wave solutions of Blandford \& Mckee.
\end{abstract}


\section{Introduction}
Self-similar solutions for fluid flows involving shock waves propagating at relativistic speeds have been presented in the literature \citep[e.g.][]{Blandford76, Hidalgo05, Nakayama05}.  Most of them are ultra-relativistic versions of the already known non-relativistic self-similar solutions. \citet{Blandford76} derived solutions for an ultra-relativistic spherical blast wave enclosed by a strong shock, which is initiated by a release of a large amount of energy within a small volume. If the released energy is not so large, the resultant non-relativistic flow is described by the well known solution discovered independently by \citet{Sedov59} and \citet{Taylor50}. The implosion of a spherical strong shock wave had been solved independently by Landau and Stanyukovich \citep[see e.g.][]{Zeldovich02} and by \citet{Guderly42} in non-relativistic cases. \citet{Hidalgo05} derived self-similar solutions for the same problem in the relativistic limit. \citet{Nakayama05} discovered solutions for the emergence of an ultra-relativistic shock wave at the surface of a star. When the shock wave is non-relativistic the same problem had been posed by \citet{Gandel'man56} and solved by \citet{Sakurai60}.

Here we present a self-similar solution for the interaction of spherical ejecta expanding at ultra-relativistic speeds with the ambient medium. When the ejecta expand at non-relativistic speeds, self-similar solutions are known to exist \citep{Chevalier82}. He discovered solutions for the flows resulting from a collision of a freely expanding spherical matter with the ambient medium. In the relativistic limit, self-similar solutions describing the shocked ambient media with power law density distributions were already obtained by \citet{Blandford76} with the shock Lorentz factor evolving  with time $t$ as $t^{-m/2}$. The constant $m$ can be determined when the nature of explosion is specified. This paper discusses self-similar solutions to describe shocked ejecta as a result of a collision with the ambient medium. This solution will be relevant to extremely high energy supernovae originated from compact progenitors with small masses. 

The density distribution of the ejecta used here is suggested by analytical solutions \citep{Nakayama05} for the breakout of an ultra-relativistic shock wave from a hydrostatic atmosphere and its non-relativistic versions \citep{Sakurai60, Matzner99}. The outer layers of the resultant flow in a non-relativistic solution expand at velocities greater than the speed of light. They have a density distribution that agrees with numerical calculations for the same problem using a special relativistic hydrodynamical code \citep{Nakamura04}. Thus we use the results as the inner boundary conditions of the present self-similar solutions. We will describe the relation of our solution with results obtained by \citet{Blandford76} in some detail.

In the next section, the density distribution of freely expanding ejecta as a result of shock breakout is discussed. 
Self-similar flows in the shocked ejecta and ambient medium are obtained in \S\ref{fs}. The behavior of these two flows at the contact surface is discussed in \S\ref{fcs}. Using the obtained solution, the shock propagation in the ambient medium is discussed in \S\ref{sprop}. The results are summarized in \S\ref{summary}.

\section{Relativistic ejecta from shock breakout} 
The velocity $v_{\rm e}$ of freely expanding ejecta at a distance $r$ from the explosion site is written as $v_{\rm e} = r/t$ at a time $t$ after the explosion. The corresponding Lorentz factor is $\gamma_{\rm e} = 1/\sqrt{1-v_{\rm e}^2}$. We set the speed of light equal to unity throughout this paper.  In the relativistic limit,  self-similar solutions for the emergence of a shock wave at the stellar surface \citep{Nakayama05} suggest  that the density in the outer layer of ejecta has a power law distribution in terms of the Lorentz factor as,
\begin{equation}\label{rhoej}
\rho_{\rm e}=\frac{\rho_{\rm e}^\prime}{\gamma_{\rm e}}=\frac{a}{t^3\gamma_{\rm e}^n},
\end{equation}
where $\rho_{\rm e}$ is the density of ejecta measured in the co-moving frame and $\rho_{\rm e}^\prime$ in the fixed frame, that is,  $\rho_{\rm e}^\prime = \rho_{\rm e} \gamma$ (hereafter densities denoted with primes mean that they are measured in the fixed frame). Here the non-dimensional parameter $n$ can be estimated to be $\sim 1.1$ if the stellar envelope is in radiative equilibrium and the adiabatic index is $4/3$. 
The constant $a$ has the dimension of a mass. On the other hand, a self-similar solution for the non-relativistic shock breakout in a plane-parallel geometry \citep{Sakurai60, Fields02} results in the density distribution in the same form with $n\sim2.6$. Though this self-similar solution assumes non-relativistic flow, the resultant density distribution agrees with that obtained by numerical calculations solving special relativistic hydrodynamics equations \citep{Nakamura04} even for large Lorentz factors $\gamma_{\rm e}>10$. The transition between these solutions occurs when the pressure to the mass density $P/\rho$ at the shock front becomes of the order of unity. Thus the outermost layers should be governed by the solution in the relativistic limit.
 
\section{Flow in the shocked region}\label{fs}
Flows  located between two shock fronts are considered by connecting two self-similar solutions. Suppose that the shock front in the ambient medium propagates at a speed $V_1$ and the shock front in the ejecta at $V_2$. If we define $\alpha = \hat{\gamma}/(\hat{\gamma}-1)$, where $\hat{\gamma}$ is the adiabatic index, the flow in each of the shocked regions is described by the following equations.
\begin{equation}\label{rhydro}
\left(\alpha P+\rho \right)\frac{d\ln \gamma}{dt}+\frac{dP}{dt}= \gamma^{-2}\frac{\partial P}{\partial t}, 
\end{equation}
\begin{equation}
\frac{d\ln\left(P^{\alpha -1}\gamma^\alpha\right)}{dt} = -\frac{\alpha}{r^2}\frac{\partial r^2v}{\partial r}, 
\end{equation}
\begin{equation}\label{rcont}
\frac{\partial\rho^\prime}{\partial t} + \frac{1}{r^2}\frac{\partial r^2\rho^\prime v}{\partial r} = 0, 
\end{equation}
where the pressure is denoted by $P$, the density $\rho$, the velocity $v$, the corresponding Lorentz factor $\gamma$, and the density in the fixed frame $\rho^\prime = \rho \gamma$. 
In the shocked ambient medium,  the ultra-relativistic equation of state is assumed. Thus $\hat{\gamma} = 4/3$ and the energy density $\epsilon$ per unit mass is written as $\epsilon=3P/\rho$ by neglecting the rest mass energy density. These approximations lead to the same equations as presented in \citet{Blandford76}. On the other hand, it is found that the rest mass energy should not be ignored in the shocked ejecta to have a compression shock. Otherwise the jump condition would result in an expansion shock for $m<1$. Then the energy density becomes $\epsilon = P/\left\{(\hat{\gamma}-1)\rho\right\} + 1$ in this region and the term including the density in the left hand side of equation (\ref{rhydro}) is not neglected.

\subsection{Jump conditions at the shock front in the ejecta}
The density, pressure, and velocity have discontinuities at the shock front. Relations to connect these quantities at both sides of the shock front are given here. The mass flux $\rho u^\mu$ and the energy momentum tensor $T^{\mu\nu}$ change across the shock propagating at the velocity of $V_2$ according to the following formulae,
\begin{eqnarray}
&[\rho u^\mu]n_\mu = 0, &\label{density} \\
&[T^{\mu\nu}]n_\nu = 0,& \label{em}
\end{eqnarray}
where $[F]\equiv  F_{\rm e} - F_2$. The subscript e refers to values in the freely expanding ejecta at radius $R_2$ defined as 
\begin{equation}
R_2(t) = \int_0^td\tau V_2(\tau),
\end{equation}
while the subscript 2 refers to values in the shocked matter at the same radius $R_2$. The variables $n_\mu$, $u^\mu$, and $T^{\mu\nu}$ have been defined as
\begin{eqnarray}
&n_\mu\equiv \Gamma_2(-V_2,\,1,\,0,\,0),\, u^\mu\equiv\gamma(1,\,v,\,0,\,0),& \\
&T^{\mu\nu}\equiv\rho hu^\mu u^\nu + P\eta^{\mu\nu}, &
\end{eqnarray}
with the metric $\eta^{\mu\nu}\equiv{\rm diag}(-1,\,1,\,1,\,1)$.
Here $\Gamma_2$ denotes the Lorentz factor of the shock and is assumed to evolve as a function of time $t$ as 
\begin{equation}
\Gamma_2^2=At^{-m}
\end{equation}
with constants $A$ and $m$. The enthalpy $h$ is denoted by other thermodynamic variables as $h\equiv\epsilon+P/\rho$.
Equation (\ref{density}) yields
\begin{equation}\label{sc1}
\rho_{\rm e}\gamma_{\rm e}(v_{\rm e}-V_2) = \rho_2\gamma_2(v_2-V_2),
\end{equation}
Equations (\ref{em}) can be rewritten as
\begin{equation}\label{sc2}
\rho_{\rm e}\gamma_{\rm e}^2(v_{\rm e}-V_2)= \rho_2h_2\gamma_2^2(v_2-V_2)+P_2V_2, 
\end{equation}
\begin{equation}\label{sc3}
\rho_{\rm e}\gamma_{\rm e}^2v_{\rm e}(v_{\rm e}-V_2) = \rho_2h_2\gamma_2^2v_2(v_2-V_2)+P_2.
\end{equation}
Equations (\ref{sc1})-(\ref{sc3}) together with the relation $\gamma_{\rm e} = \sqrt{m+1} \Gamma_2$ at $r = R_2$ lead to the following equation
\begin{equation}\label{eq-q}
\alpha x^3 + (\alpha -2)\sqrt{m+1} x^2 - (\alpha -2)x -\alpha \sqrt{m+1}=0,
\end{equation}
where $x = \gamma_2 / \Gamma_2$. This equation has the only one positive solution between 1 and $\sqrt{m+1}$ for $m>-1$. If this solution is denoted as $x=\sqrt{q}$,  the Lorentz factor, density, and pressure at the front in the shocked matter will be derived as
\begin{equation}
\gamma_2^2 = q\Gamma_2^2,
\end{equation}
\begin{equation}
\rho^\prime_2 = \frac{mq}{(m+1)(q-1)}\rho_{\rm e}\gamma_{\rm e},
\end{equation}
\begin{equation}
P_2 = \frac{m\rho_{\rm e}}{\alpha(q-1)+2} \left( 1-\sqrt{\frac{q}{m+1}} \right).
\end{equation}
These relations are used to obtain the boundary conditions for flows in the shocked ejecta. If the constant $m$ is smaller than unity as is usually the case, the compression is so weak that the thermal energy cannot dominate the rest mass energy in the shocked ejecta. Therefore we cannot use the ultra-relativistic equation of state there while the ultra-relativistic equation of state is always a good approximation in the shocked ambient medium as long as the shock Lorentz factor is much larger than unity.

\subsection{Self-similar flow in the shocked ejecta}\label{sfse}
It is convenient to define self-similar variables for the pressure $F(\xi)$, the Lorentz factor $G(\xi)$, and the density $H(\xi)$ in the shocked ejecta as
\begin{equation} \label{p@e}
P = \frac{m\rho_{\rm e}}{\alpha(q-1)+2} \left( 1-\sqrt{\frac{q}{m+1}} \right)F(\xi),
\end{equation}
\begin{equation} \label{g@e}
\gamma^2 = q\Gamma_2^2G(\xi), 
\end{equation}
\begin{equation}
\rho^\prime = \frac{mq}{(m+1)(q-1)}\rho_{\rm e}\gamma_{\rm e}H(\xi).
\end{equation}
The similarity variable $\xi$ is defined as
\begin{equation}
\xi = \left\{1+2(m+1)\Gamma_2^2\right\}\left(1-\frac{r}{t}\right),
\end{equation}
following \citet{Blandford76}.

To obtain equations governing the self-similar evolution, the transformation of coordinates from $(r,\,t)$ to $(\Gamma_2,\, \xi)$ is performed using the following relations of derivatives;
\begin{equation}
\frac{\partial}{\partial\ln t} = -m\frac{\partial}{\partial\ln\Gamma_2^2}+\left\{(m+1)(2\Gamma_2^2-\xi)+1\right\}\frac{\partial}{\partial\xi}, 
\end{equation}
\begin{equation}
t\frac{\partial}{\partial r} = -\left\{1+2(m+1)\Gamma_2^2\right\}\frac{\partial}{\partial\xi}, 
\end{equation}
\begin{equation}
\frac{d}{d\ln t} = -m\frac{\partial}{\partial\ln\Gamma_2^2}+(m+1)\left(\frac{1}{qG}-\xi\right)\frac{\partial}{\partial\xi}.
\end{equation}
 Then we obtain from equations (\ref{rhydro})--(\ref{rcont}),
\begin{equation} \label{G@e}
2(m+1)(1+qG\xi)\frac{d\ln F}{d \xi}-(m+1)(\alpha+I)(1-qG\xi)\frac{d\ln G}{d \xi}= \{mn-(\alpha +I)m -6\}qG,
\end{equation}
\begin{equation}
2(\alpha-1)(m+1)(1-qG\xi)\frac{d\ln F}{d \xi}-\alpha(m+1)(1+qG\xi)\frac{d\ln G}{d \xi}=\{(1-\alpha)mn + \alpha m +2\alpha -6\}qG,
\end{equation}
\begin{equation} \label{H@e}
2(m+1)(1-qG\xi)\frac{d\ln H}{d \xi} - 2(m+1)\frac{d\ln G}{d \xi} = -(mn-m-2)qG,
\end{equation}
where $I = I(\xi) = \rho/P$. The boundary conditions are given at the shock front as
\begin{equation}\label{bc@e}
G(1) = F(1) = H(1) = 1.
\end{equation}
The solution is obtained by numerically integrating equations (\ref{G@e})--(\ref{H@e}) from the shock front $\xi=1$ to the contact discontinuity $\xi = \xi_{\rm c}=1/qG(\xi_{\rm c})$.

\subsection{Jump conditions at the shock front in the ambient medium}
The density, Lorentz factor, and pressure change across the shock wave propagating at the speed $V_1=\sqrt{1-1/\Gamma^2}$ according to the following relations, 
\begin{equation}
\rho^\prime_1 = \frac{\rho_{\rm i}\Gamma^2}{2}, \,\gamma_1^2 = \frac{\Gamma^2}{2},\,P_1 = \frac{2\rho_{\rm i}\Gamma^2}{3},
\end{equation}
where the subscript 1 refers to values in the shocked fluid at the shock front, while the subscript i refers to values in the unshocked ambient medium at the shock front. The shock Lorentz factor is denoted by $\Gamma$. This $\Gamma$ evolves with time in the same way as $\Gamma_2$, i.e., $\Gamma^2 = B t^{-m}$. The density in the ambient medium is assumed to have a power law distribution as $\rho=br^{-k}$ where $b$ is a constant. Thus self-similar variables for the pressure $f(\chi)$, Lorentz factor $g(\chi)$, and density $h(\chi)$ are defined as
\begin{equation}
P=\frac{2\rho_{\rm i}\Gamma^2}{3}f(\chi), \label{p@a}
\end{equation}
\begin{equation}
\gamma^2 = \frac{\Gamma^2}{2}g(\chi), \label{g@a}
\end{equation}
\begin{equation}
\rho^\prime = 2\rho_{\rm i}\Gamma^2h(\chi). 
\end{equation}
Here the similarity variable $\chi$ has been introduced as
\begin{equation}
\chi=\left\{1+2(m+1)\Gamma^2\right\}\left(1-\frac{r}{t}\right).
\end{equation}

\subsection{Self-similar flow in the shocked ambient medium}
The same procedure in \S\ref{sfse} leads to equations governing a self-similar flow in the shocked ambient medium as
\begin{equation}
\frac{d\ln f}{gd\chi}=\frac{4\{2(m-1)+k\}-(m+k-4)g\chi}{(m+1)(4-8g\chi+g^2\chi^2)},  \label{f@a}
\end{equation}
\begin{equation}
\frac{d\ln g}{gd\chi}=\frac{(7m+3k-4)-(m+2)g\chi}{(m+1)(4-8g\chi+g^2\chi^2)}, 
\end{equation}
\begin{equation}\label{h@a}
\frac{d\ln h}{gd\chi}=\frac{2(9m+5k-8)-2(5m+4k-6)g\chi+(m+k-2)g^2\chi^2}{(m+1)(4-8g\chi+g^2\chi^2)(2-g\chi)},
\end{equation}
in the relativistic limit. The boundary conditions are
\begin{equation}\label{bc@a}
f(1) = g(1) = h(1) = 1.
\end{equation}
These are exactly the same equations as presented in \citet{Blandford76}.

\section{Flow at the contact surface}\label{fcs}
 The density distribution has a discontinuity at the contact surface defined as $\xi=\xi_{\rm c}$ and $\chi=\chi_{\rm c}$ where $G(\xi_{\rm c})\xi_{\rm c} = 1/q$ and $g(\chi_{\rm c})\chi_{\rm c} = 2$, while the velocity and pressure are continuous. The same velocity of the two flows at the contact surface indicates
 that the ratio of Lorentz factors of the two shock fronts should become
\begin{equation}\label{lorentz12}
\frac{\Gamma^2}{\Gamma_2^2} = \frac{B}{A} = 2q\frac{G(\xi_{\rm c})}{g(\chi_{\rm c})},
\end{equation}
according to equations (\ref{g@e}) and (\ref{g@a}). The continuous pressure at the contact surface requires that the pressures at both sides of the contact surface evolve with time in the same manner.  Thus we obtain from equations (\ref{p@e}) and  (\ref{p@a})
\begin{equation}\label{tev}
m = \frac{6-2k}{n+2}.
\end{equation}
This indicates that $k<4+n/2$ because $m>-1$. 
In addition, the pressure equilibrium yields a relation of constants that have been introduced in the preceding sections as
\begin{equation}\label{scale}
\frac{a}{bA^{1+n/2}}=\frac{4q\{ \alpha (q-1)+2\}G(\xi_{\rm c})f(\chi_{\rm c})}{3mg(\chi_{\rm c})F(\xi_{\rm c})} \left( \frac{m+1}{\xi_{\rm c}}\right)^{n/2} \left( 1- \sqrt{\frac{q}{m+1}} \right)^{-1}.
\end{equation}
This relation will be used to discuss the propagation of the outer shock in the ambient medium.

A blast wave in a uniform ambient medium with power supply was treated by \citet{Blandford76}. They assumed that the power is supplied by a stationary source located at $r=0$ and varies with time as a power law, $L=L_0t^s$. If $s = (2n-8)/(n+2)$, then the flow in the shocked ambient medium evolves exactly in the same way as described in \citet{Blandford76}.  They also argued that  the adiabatic impulsive solution is appropriate for the flow with $m$ in the range of $m>3$ or $m<-1$. Since $m>-1$ in our problem, the condition $m>3$ restricts the value of $n$ to be less than $-2k/3$ through equation (\ref{tev}). Thus the values of $n=$ 1.1 and 2.6 indicate a continuous supply of energy.

When $(n,k) = (1.1,\,2)$, equation (\ref{tev}) yields $m=2/(n+2)\sim0.65$ and then, from equation (\ref{eq-q}), $q\sim1.06$. Performing numerical integration of equations (\ref{G@e})--(\ref{H@e}) and (\ref{f@a})--(\ref{h@a}) with these parameters, boundary conditions (Eqs. (\ref{bc@e}) and (\ref{bc@a})), and $\hat{\gamma}=5/3$ in the shocked ejecta, we find $\xi_{\rm c}\sim0.95$ and $(f_{\rm c}, \,g_{\rm c}, \, F_{\rm c},\, G_{\rm c}) \sim (0.18,\, 0.47,\, 2.1,\, 0.98)$. Here the subscript c of each variable indicates that the value is taken at the contact surface. Results are shown in Figure 1a. 
When $(n,k) = (2.6,\,2)$, the same procedure leads to $m=2/(n+2)\sim0.43$, $q\sim1.05$, $\xi_{\rm c}\sim0.97$, and $(f_{\rm c}, \,g_{\rm c}, \, F_{\rm c},\, G_{\rm c}) \sim (0.32,\, 0.70,\, 2.0,\, 0.98)$. Results are shown in Figure 1b.

\section{Propagation of  shock in  stellar wind}\label{sprop}
\citet{Nakayama05} investigated the evolution of an ultra-relativistic shock wave in a plane-parallel atmosphere and derived self-similar solutions. The resultant energy spectrum of the ejected matter with an explosion energy $E_{\rm ex}$ and an ejected mass $M_{\rm ej}$ for a radiative stellar envelope can be expressed as 
\begin{equation}\label{Edistr2}
\frac{M(>\varepsilon)}{M_{\rm ej}} \propto \left( \frac{E_{\rm ex}}{M_{\rm ej}} \right)^{5.75} \varepsilon^{-2.10},
\end{equation}
where $M(>\varepsilon)$ denotes the ejected mass that have particle energy per nucleon greater than $\varepsilon = m_p (\gamma_{\rm e} -1)$. Here the proportionality constant is not specified by \citet{Nakayama05} because their solutions cannot trace the flow to the free expansion phase in which the equation of state is no longer ultra-relativistic. 

Using the relation $dM = 4 \pi r^2 \rho_{\rm e} \gamma_{\rm e} dr$ we can convert the energy distribution given by equation (\ref{Edistr2}) into the density distribution of the ejecta as a function of
$\gamma_{\rm e}$ and $t$.
\begin{equation}\label{eqb_01}
\rho_{\rm e} = a  \gamma_{\rm e}^{-n} t^{-3} \propto \left(\frac{E_{\rm ex}}{M_{\rm ej}}\right)^{5.75}  M_{\rm ej} \gamma_{\rm e}^{-1.10} t^{-3}.
\end{equation}
This formula indicates $a \propto E_{\rm ex}^{5.75} M_{\rm ej}^{-4.75}$ in equation (\ref{rhoej}), $n=1.10$, and $m=0.645$ from equation (\ref{tev}). Eliminating $A$ in equations (\ref{lorentz12}) and (\ref{scale}) leads to
\begin{equation}\label{L2}
\Gamma^2 = \left[ \frac{3m(2q)^{n/2}}{2\{ \alpha (q-1)+2\}}\left(\frac{\xi_{\rm c}}{m+1} \right)^{n/2} \left( 1- \sqrt{\frac{q}{m+1}}\right) \frac{F_{\rm c} G_{\rm c}^{n/2}}{f_{\rm c} g_{\rm c}^{n/2}} \frac{a}{b} \right]^{2/(n+2)} t^{-m}.
\end{equation}
Substituting the expression for $a$ into equation (\ref{L2}) with the exponents $m$ and $n$ obtained above, then we have
\begin{equation}\label{L3}
\Gamma \propto \left(a/b\right)^{1/(n+2)} t^{-m/2} \propto E_{\rm ex}^{1.85} M_{\rm ej}^{-1.53} t^{-0.322}.
\end{equation}

We can discuss the time evolution of the Lorentz factor of a shock in a steady stellar wind more definitely if the proportionality constant in equation  (\ref{Edistr2}) is specified. This is the case for  Type Ic supernovae (SNe Ic) discussed by \citet{Fields02} and \citet{Nakamura04}.
According to the results of special-relativistic hydrodynamical calculations by \citet{Nakamura04} \citep[see also][]{Fields02}, the energy distribution of SN Ic ejecta is given by
\begin{equation}\label{Edistr}
\frac{M(>\varepsilon)}{M_{\rm ej}}=
C \left(\frac{E_{\rm ex}}{10^{51}\,\mbox{ergs}}\right)^{3.4}\left(\frac{M_{\rm ej}}{1\, \Msun}\right)^{-3.4}\left(\frac{\varepsilon}{10\,\mbox{MeV}}\right)^{-3.6},
\end{equation}
where $C$ is a non-dimensional constant ($\sim 2\times10^{-4}$) slightly  depending on stellar models.  
Thus the density distribution of the ejecta becomes
\begin{equation}\label{eqb_02}
\rho_{\rm e} =  2.26 \times 10^{-8} C M_{\rm ej} \left(\frac{E_{\rm ex}}{10^{51}\,\mbox{ergs}}\right)^{3.4} \left(\frac{M_{\rm ej}}{1\, \Msun}\right)^{-3.4} \gamma_{\rm e}^{-2.6} t^{-3}.
\end{equation}
Accordingly the constant $a$ is expressed in terms of the parameters of explosion as
\begin{equation}\label{constantaej}
a =2.26 \times 10^{-8} C M_{\rm ej}\left(\frac{E_{\rm ex}}{10^{51}\,\mbox{ergs}}\right)^{3.4}\left(\frac{M_{\rm ej}}{1\, \Msun}\right)^{-3.4}.
\end{equation}
On the other hand, the constant $b$ is expressed in terms of the mass loss rate $\dot{M}$ and the wind velocity $v_{\rm wind}$ as $b=\dot{M}/4\pi v_{\rm wind}$.  Substituting the expression for $a$ obtained above and values for non-dimensional functions at the contact surface into equation (\ref{L2}),  
$\Gamma$ is expressed in terms of physical quantities characterizing the explosion and ambient medium as
\begin{equation}\label{LoutSN}
\Gamma = 6.3 \times 10^{1} \left( \frac{b}{10^{10}{\rm \,g\,cm}^{-1}}\right)^{-0.22} \left( \frac{E_
{\rm ex}}{10^{53}\,\mbox{ergs}}\right)^{0.74}\left(\frac{M_{\rm ej}}{1\, \Msun}\right)^{-0.52}\left(\frac{t}{1{\rm \,sec}}\right)^{-0.22}.
\end{equation}

\section{Summary}\label{summary}
We have found self-similar solutions for the collision of spherical ejecta freely expanding at relativistic speeds with the ambient medium when the ejecta have a density distribution in the form of equation (\ref{rhoej}) and the ambient medium has a power law density distribution in terms of the distance from the explosion site with the exponent $-k$. Solutions can be obtained as long as $k<4+n/2$. It is found that the rest mass energy is always comparable to the thermal energy in the shocked ejecta while the rest mass energy becomes negligible in the shocked ambient medium.

Profiles of the flows in the shocked regions are presented in two cases where the ejecta are results of the shock breakout from a stellar surface embedded in the steady wind. The time evolution of the Lorentz factor of the outer shock front is derived as a function of physical parameters characterizing the explosion and the environment. 

We thank an anonymous referee for important remarks.
This work has been partially supported by the grant in aid (16540213) of the Ministry of Education, Science, Culture, and Sports in Japan.

\clearpage
\begin{figure}
\begin{center}
\epsscale{.80}
\plotone{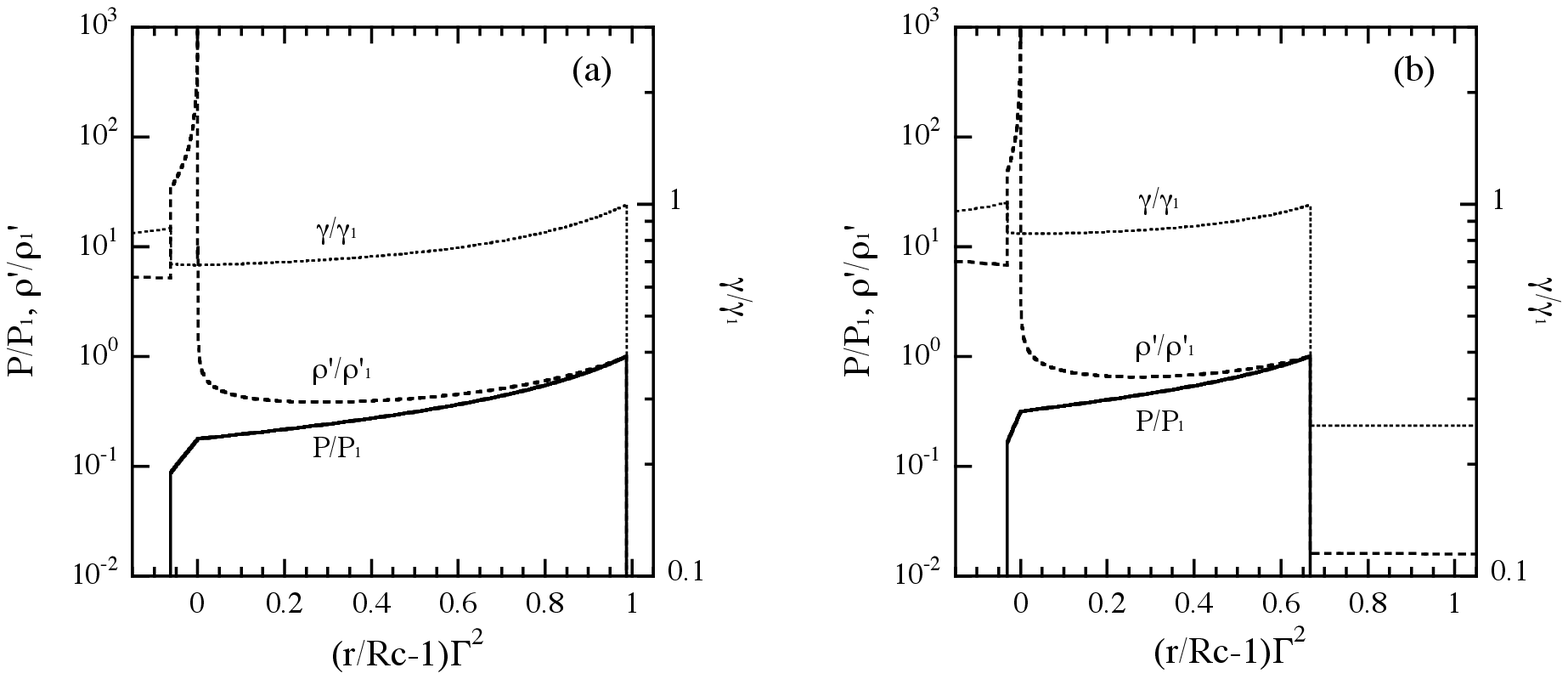}
\caption{(a): Profiles of the pressure $P$ (solid line), the density $\rho$ (dashed), and the gas Lorentz factor $\gamma$ (dash-dotted) in the interaction region between freely expanding ejecta with $n=1.10$ and the ambient medium with a power low distribution ($\rho \varpropto r^{-2}$). The physical variables are normalized to their values at $r=R_1$ where the subscript $1$ refers to the value at the outer shock front.  The radius at the contact surface is denoted by $R_{\rm c}$. (b): Profiles of the same quantities as (a) but  for freely expanding ejecta ($E_{\rm ex}=10^{53}$ ergs and $M_{\rm ej}=13\, \Msun$) at time $t=100$ sec. }
\label{fig1}
\end{center}
\end{figure}
\clearpage

\end{document}